# Comparison Between PBE-D3, B3LYP, B3LYP-D3 and MP2 Methods for Quantum Mechanical Calculations of Polarizability and IR-NMR Spectra in $C_{24}$ Isomers, Including a Novel Isomer with $D_{2d}$ Symmetry

Futtaim Alhanzal [a, *, d] , Nabil Joudieh [a, *, d], Khansaa Hussein [c,*,d], and Nidal Chamoun [b,*,d]

[a] Faculty of sciences, Physics Department, Damascus University, P.O. Box 30621, Damascus, Syria.
[b] Physics Department, HIAST, P.O. Box 31983, Damascus, Syria .
[c] Faculty of sciences, Chemistry  Department, Damascus University, P.O. Box 30621, Damascus, Syria.
* Correspondence: fattomm90@gmail.com   ; futtaim91.alhandel@damascusuniversity.edu.sy  ; njoudieh@yahoo.fr ;  khansaa1.hussein@damascusuniversity.edu.sy;  nidal.chamoun@hiast.edu.sy.
[d] These authors contributed equally to this work.

**Abstract**

In this study, we performed a comprehensive theoretical analysis of $C_{24}$ isomers in the gaseous phase using the PBE-D3/cc-pVTZ, B3LYP/cc-pVTZ, B3LYP-D3/cc-pVTZ and MP2/6-31G methods. We also considered the basis set cc-pVTZ for the MP2 method, and carried out optimized (single point) calculations in three (remaining two, where convergence was too much time consuming) isomers in order to reveal the potentiality of the method. Our investigation covered a wide range of properties, including geometry optimizations, chemical stability, polarizabilities, nuclear screening constants, Fermi (FE), gap (GE), and atomization energies (AE), thermodynamic analysis, reactivity index, as well as IR and NMR spectra. These calculations were performed for the ring ($D_{12h}$), sheet ($D_{6h}$) and two cage ($D_{6d}$ and $O_h$) configurations. Interestingly, we also proposed a new structure, the bracelet ($D_{2d}$) arrangement, which appeared to be stable according to the PBE, B3LYP and B3LYP-D3 methods, but was classified as a transition state by the MP2 method. The results consistently indicated that the $D_{6h}$ isomer is the most stable one among the $C_{24}$ isomers studied, while the $D_{2d}$ isomer was found to be the least stable. Regarding the gap energy (GE), the B3LYP and B3LYP-D3 methods consistently yielded higher values compared to the PBE's, with an average DFT (PBE and B3LYP) GE of 1.89 eV, whereas the MP2 method showed a substantially higher GE value of 7.6 eV, representing an increase of approximately 75%. Additionally, the polarizabilities of the $C_{24}$ isomers were found to be overestimated by the PBE, B3LYP, and B3LYP-D3 methods when compared to the corresponding MP2 values. The PBE-D3 method consistently produces higher polarizabilities for the $C_{24}$ isomers in comparison to B3LYP, B3LYP-D3, and MP2 methods. The investigation confirms that the $O_h$ ($D_{12h}$) isomer has the smallest (largest) polarizability, as agreed upon by all methods. Moreover, the polarizability of $D_{12h}$ is notably affected by the selected DFT method, while that of $O_h$ displays lower sensitivity but shares similarities with $D_{6d}$. However, for the newly proposed $D_{2d}$ isomer, the polarizability is ranked third (fourth) in ascending order with the PBE-D3 (B3LYP-D3) method. This highlights the importance of considering the electronic correlation and dispersion effects in accurately predicting polarizabilities. The results obtained from different methods shed light on the impact of methodology choice on the predicted properties, emphasizing the need for careful consideration when analyzing and interpreting theoretical results for such various geometries.

**Graphical Abstract**

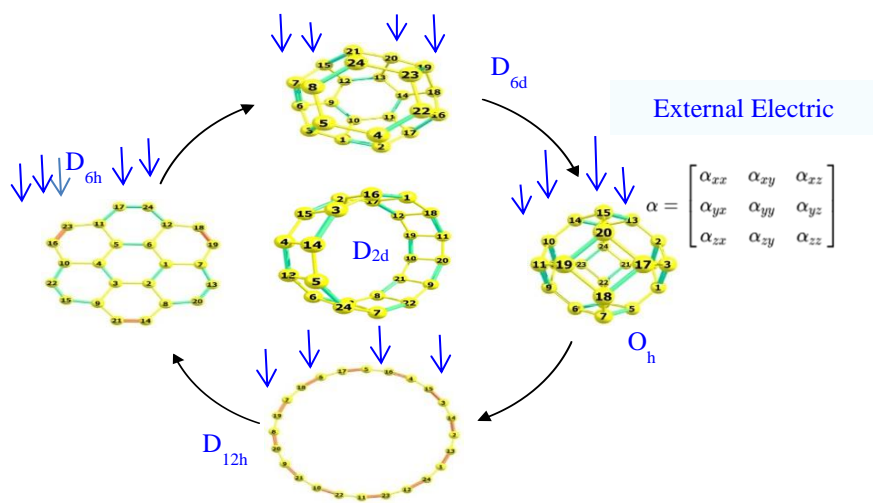

Contents

# 1. Introduction

Fullerenes, which were discovered in the late eighties [1], are a fascinating form of carbon, known as its third allotropic form. Fullerenes consist of sp$^2$ hybridized carbon atoms bonded together in the form of a hollow sphere with different sizes such as $C_{20}$, $C_{24}$, $C_{60}$, etc. [2]. Among the various fullerenes, $C_{20}$ is the most recently discovered one [3]. Subsequently, medium and large sizes of fullerenes molecules were discovered, with some reaching up to 6000 carbon atoms [4]. Following the discovery of the $C_{60}$ fullerene in 1985 [5], numerous theoretical and experimental studies have been conducted aiming at investigating the structures and stability of small fullerenes containing fewer than 60 carbon atoms, by using the mass spectrometry methods [6-16]. The C24 fullerene, one of the smallest fullerenes, was first reported in 1993 [17] and produced, under certain experimental conditions, from carbon vapor condensation. It has four isomers, namely the ring, sheet, and cages (fullerene-like), but the most common is the $D_{6d}$ cage isomer which contains two hexagons with 12 pentagons between them [6,18,19]. At the theoretical level, Jensen and Toftlund [20] investigated the four isomers of $C_{24}$ fullerene: cages ($D_{6d}$, $O_h$), ring ($D_{12h}$), and sheet ($D_{6h}$) using the Hartree-Fock (HF) method and second-order Moller-Plesset perturbation theory (MP2) with a DZP basis set. Their study showed that the $D_{6h}$ sheet isomer is the most stable, followed by the $D_{6d}$ cage isomer. In 1998 [21], the energy difference between the ring and fullerene forms of $C_{24}$ has been calculated using ab initio methods, which were compared to density functional methods. The calculations strongly suggest that the fullerene form is favored by 80 kcal/mol over a monocyclic ring structure. In 1997, Balevišius *et al.* [22] studied the chemical stability and electronic properties of the $D_{6d}$ isomer using the PM3 method implemented in the MOPCA program [23]. In 2007, the geometrical parameters, total energy, heat of formation, energies of HOMO and LUMO orbitals, and density of one electron states (DOS) were determined by using semi-empirical quantum chemistry PM3 method for cubic polymerized structures of the $O_h$ isomer [24]. The results of calculations allow for the existence of a polymerized cubic crystal structure based on all the considered small fullerenes. In 2012, Anafche and Naderi [25] reported results concerning the structural stabilities, geometry, electronic properties, and binding energies of $C_{24}$ ($D_{6d}$) and some of its hetero fullerene derivatives at the B3LYP/6-311-EFG**//B3LYP/6-31+G* level of theory. Recently, several studies [26-28] focused on the $C_{24}$ fullerene and its four isomers by studying their geometry, energy stability, spectra, and interactions with other molecules. Using density functional theory (DFT) and coupled cluster calculations, the relative energies and infrared spectra have been determined for the four different types of $C_{24}$ isomers. Among the four isomers, it was found that the graphene (sheet) form of $C_{24}$ best accommodates astronomical data [28]. Recent research activities in various fields including molecular electronics, molecular devices, nanometer electronics, nanotechnology, energy storage, and biomedical/nanomedicine applications, demonstrate that fullerene $C_{24}$ is becoming increasingly connected to several nano themes. As a carbon material, the $C_{24}$ isomer is considered a promising candidate for future developments in nanotechnology, for both civil and military/ defense applications. Studies have demonstrated its potential in fields such as superconductivity and electronic transport properties, highlighting its applicability in nanometer electronics. The importance of $C_{24}$ is also seen in its hydrogen storage capabilities, an important consideration in the energetic domain [29,30]. An important view on these directions is the work of Sawhney et al. [29] regarding the superconductive properties of $C_{24}$ and its doped counterparts. Their results clearly exhibited the better electrical performance of the pure $C_{24}$ over its doped counterparts for cryogenic electronic applications. As to single molecular devices and nanometer electronics, we report the work of When-Kai Zhao et al. [30], which studied the orientation effect on the electronic transport properties of $C_{24}$ fullerene between the electrodes (Au-$C_{24}$-Au). These results confirm the applicability potential and importance of $C_{24}$ fullerene in nanometer electronics. The energy storage domain constitutes another important application. Actually, the hydrogen storage capability in fullerene constitutes an increasingly significant area of interest, and the hydrogen storage properties of $C_{24}$ fullerene were studied using the DFT recently [31-33], where this capacity was found to approximate 10-12 wt. %. Also, Mahamiya et al. [34] conducted a theoretical study on the hydrogen storage capacity of yttrium atom-decorated $C_{24}$ fullerene. Their results show that a single yttrium atom attached to $C_{24}$ fullerene can reversibly adsorb a maximum number of 6 $H_2$ molecules. Using the DFT, Mahamiya et al. [35] showed in 2022, that the scandium-decorated $C_{24}$ fullerene can adsorb up to six hydrogen molecules with an average adsorption energy of −0.35 eV per $H_2$ and an average desorption temperature of 451 K, and also demonstrated that the scandium-decorated $C_{24}$ fullerene system is thermodynamically stable, providing thus a potential promising candidate for a reversible high-capacity hydrogen storage device.

In this study, we focus on the $C_{24}$ fullerene and its importance in various fields such as technology and nano-science. Since there is relatively limited theoretical research on $C_{24}$, we aim to bring a contribution by examining its isomers. In our investigation, based on the study conducted by Zhang and Dolg [36], we have introduced a novel structure called the "bracelet" with $D_{2d}$ symmetry and thoroughly examined its stability using computational methods. To assess stability, we employed the PBE-D3, B3LYP and B3LYP-D3 methods with the cc-pVTZ basis set and the MP2 method with two basis sets 6-31G and cc-pVTZ. By calculating the IR spectra frequencies, we were able to confirm the stability of the bracelet structure using the PBE-D3, B3LYP and B3LYP-D3 methods with the cc-pVTZ basis set. However, when we applied the Møller-Plesset second-order perturbation theory (MP2/6-31G), we encountered a discrepancy indicating that the structure may exhibit characteristics of a transition state. This inconsistency is likely due to the utilization of a non-extended basis set (6-31G) during the optimization process for the MP2 calculations, and to the fact that the latter method in itself does not take the dispersion interactions into consideration, in contrast to the PBE method which accounts explicitly for the dispersion, and the B3LYP which does this implicitly through suitable parametrization.

Unsatisfactorily, the MP2/cc-pVTZ optimization for the $D_{2d}$ (Bracelet) and $D_{6d}$ (Cage) isomers did not converge, even after an unusually long computation time. The fact that the optimization did converge for the $O_h$ (cage), the $D_{6h}$ (sheet), and the $D_{12h}$ (ring) isomers suggests that the structural arrangement in a cage (bracelet) for $D_{6d}$ ($D_{2d}$) may not be the primary cause of the convergence issue. It is possible that the intrinsic structure of the potential energy hypersurface, combined with the nature of the MP2 method and the size of the basis set, is the underlying reason for the convergence problem. In order to qualitatively assess the impact of extending the basis set from 6-31G to cc-pVTZ using the MP2 method, a single MP2/cc-pVTZ point calculation was performed for the studied isomers.

Furthermore, we conducted a comprehensive quantum mechanical analysis on classical fullerenes and nonfullerene isomers. This analysis encompassed the use of various computational methods, including PBE-D3, B3LYP, B3LYP-D3 and MP2, implemented through the ORCA 5.0.1 program package. We thoroughly investigated several properties such as static electric polarizabilities, NMR and IR spectra, energy quantities (including gap, Fermi, and atomization energies), as well as thermodynamic properties. The results obtained were meticulously examined and presented in the study.

## 2. Computational Methods

The computational calculations were carried out using the ORCA 5.0.1 program package [37]. Four distinct levels of calculations were employed: PBE-D3, B3LYP, B3LYP-D3 with the cc-pVTZ basis set, and MP2/6-31G. For the PBE-D3, B3LYP and B3LYP-D3 calculations, the optimization of molecular geometry was performed along with the determination of energetic properties, static electric Polarizabilities, dipole moment, and the computation of IR frequencies as well as IR and NMR spectra. The same perturbation order in MP2/6-31G was utilized to calculate these parameters. An optimized (single) energy calculation using MP2/6-31G (MP2/cc-pVTZ) was also carried out. Chemcraft [38], a graphical interface for drawing, was used in conjunction with the ChemDraw feature to obtain the visual representations of the five shapes and to generate the corresponding IR and NMR spectra.

## 3. $C_{24}$ Isomers: Quantum Mechanical Characterization and Properties

A comprehensive quantum mechanical analysis was conducted to characterize the structures and properties of the five $C_{24}$ isomers ($D_{6d}$, $O_h$, $D_{12h}$, $D_{6h}$, and $D_{2d}$) using four distinct levels of theory. The analysis encompassed tasks such as geometry optimization, evaluation of chemical stability, calculation of polarizabilities and various energy parameters, determination of vibration frequencies, and theoretical predictions of both infrared and NMR spectra for each isomer at all four levels of theory, whereas the optimized energies and thermodynamic properties of the isomers under study are reported in Tables S1 and S2 of the supporting information file.

### 3.1. Geometry optimizations of $C_{24}$ isomers

Table 1 to 5 present the optimized key bond lengths and angles obtained from calculations using the PBE-D3/cc-pVTZ, B3LYP/cc-pVTZ, B3LYP-D3/cc-pVTZ and MP2/6-31G methods for the five $C_{24}$ isomers, and from the MP2/cc-pVTZ method for the $D_{12h}$, $D_{6h}$ and $O_h$ isomers.

Bond lengths are denoted as r (in Angstroms), while the angles are denoted as A (in degrees). The optimized geometries of the five $C_{24}$ isomers are depicted in Figure 1. The $D_{6d}$ cage structure consists of one hexagonal ring at the top and bottom, with 12 pentagonal rings in the middle. A total of 36 C-C bonds in the $D_{6d}$ structure are classified into four categories, whereas in the $O_h$ cage structure these bonds alternate between single and double character. The optimized bond lengths are in good agreement with previously reported values of 1.38 Å and 1.50 Å [39]. The sheet isomer $D_{6h}$ comprises interconnected hexagonal polygons, while the ring isomer $D_{12h}$ features a circular arrangement of carbon atoms. The newly proposed $D_{2d}$ isomer consists of two rings connected by single bonds, along with 12 quadrilateral polygons in the middle. Figure 1 illustrates the optimized structures obtained using the PBE-D3/cc-pVTZ method for the $C_{24}$ isomers.

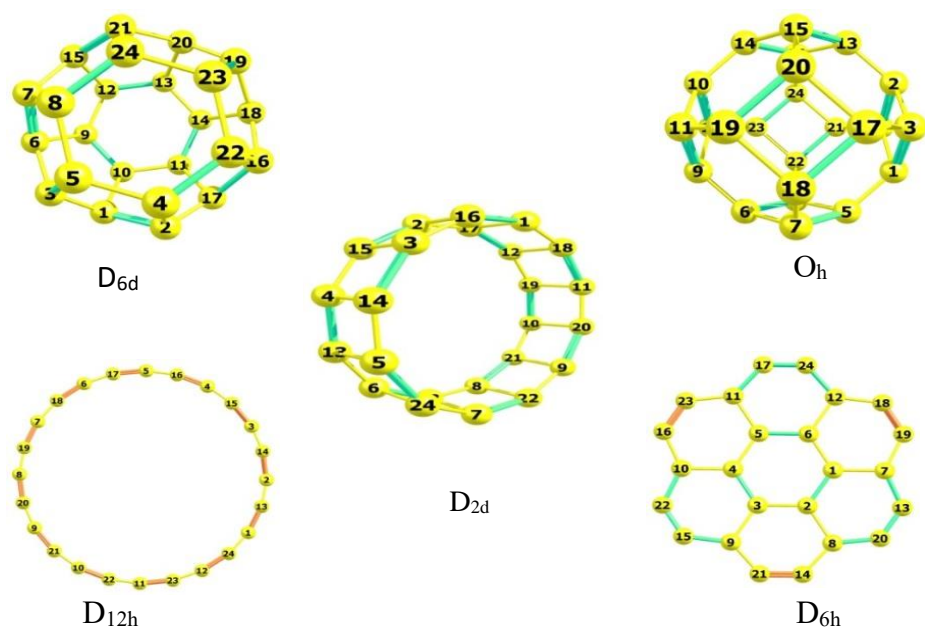

**Fig.1.** Optimized PBE-D3/cc-pVTZ geometry of the five $C_{24}$ isomers.

**Table 1.** Bond lengths (r in Angstrom) and angles' measurements (A in degrees) at PBE-D3/cc-pVTZ, B3LYP/cc-pVTZ, B3LYP-D3/cc-pVTZ, MP2/cc-pVTZ and MP2/6-31G levels for the $D_{6d}$ Cage.

| Isomer | $D_{6d}$ Cage | | | | | | |
|---|---|---|---|---|---|---|---|
| Basis set | cc-pVTZ | | | | 6-31G | B3LYP/ 6-31G(d) [27] | MP2/DH(d) [21] |
| Method | PBE-D3 | B3LYP | B3LYP-D3 | MP2 | MP2 | | |
| r(19-18) | 1.465 | 1.458 | 1.462 | - | 1.405 | 1.398 | 1.365 |
| r(19-20) | 1.428 | 1.358 | 1.360 | - | 1.476 | 1.457 | 1.463 |
| r(19-23) | 1.443 | 1.529 | 1.535 | - | 1.556 | 1.523 | 1.531 |
| r(22-23) | 1.465 | 1.417 | 1.418 | - | 1.450 | 1.437 | 1.423 |
| r(10-11) | 1.409 | - | - | - | - | - | - |
| r(21-24) | 1.528 | - | - | - | - | - | - |
| r(22-16) | 1.512 | - | - | - | - | - | - |
| A(2-1-10) | 106.58 | 107.34 | 107.19 | - | 108.93 | - | - |
| A(2-1-3) | 108.85 | 109.10 | 109.22 | - | 109.11 | - | - |
| A(6-3-5) | 105.87 | 107.30 | 107.16 | - | - | - | - |
| A(5-4-22) | 118.19 | 119.94 | 119.94 | - | 119.99 | - | - |
| A(6-9-10) | 108.37 | 106.05 | 106.12 | - | - | - | - |
| A(9-10-11) | 119.19 | 120.00 | 120.01 | - | - | - | - |
| A(1-3-5) | 109.86 | - | - | - | 107.55 | - | - |
| A(1-3-6) | 112.70 | - | - | - | - | - | - |
| A(3-1-10) | 103.16 | - | - | - | 107.56 | - | - |
| A(2-4-22) | 107.39 | - | - | - | 106.25 | - | - |
| A(4-5-8) | 122.61 | - | - | - | 120.00 | - | - |
| A(1-2-17) | 110.24 | - | - | - | - | - | - |

**Table 2**. Bond lengths (r in Angstrom) and angles' measurements (A in degrees) at PBE-D3/cc-pVTZ, B3LYP/cc-pVTZ, B3LYP-D3/cc-pVTZ, MP2/cc-pVTZ and MP2/6-31G levels for the $O_h$ Cage.

| Isomer | $O_h$ Cage | | | | | | |
|---|---|---|---|---|---|---|---|
| Basis set | cc-pVTZ | | | | 6-31G | B3LYP/ 6-31G(d) [27] | HF/DZP [20] |
| Methods | PBE-D3 | B3LYP | B3LYP-D3 | MP2 | MP2 | | |
| r(2,3) | 1.494 | 1.489 | 1.494 | 1.483 | 1.520 | 1.457 | 1.463 |
| r(2,13) | 1.378 | 1.368 | 1.368 | 1.387 | 1.402 | 1.398 | 1.365 |
| A(20-17-18) | 90.00 | 90.00 | 90.00 | 90.00 | 90.00 | - | - |
| A(13-2-3) | 120.00 | 120.00 | 120.00 | 120.00 | 120.00 | - | - |

Tables 1 and 2 provide the bond lengths and angles for the $D_{6d}$ and $O_h$ cage at the PBE-D3/cc-pVTZ, B3LYP/cc-pVTZ, B3LYP-D3/cc-pVTZ, and MP2/6-31G levels. The $D_{6d}$ cage exhibits four types of carbon-carbon bond lengths at the MP2/6-31G level, with an average length of 1.477 Å. In contrast, the $O_h$ isomer shows two types of bond lengths across all four methods, including the MP2 method in two different basis sets. The B3LYP/cc-pVTZ and PBE-D3/cc-pVTZ methods yield similar results, with an increase in bond types observed for the $D_{6d}$ cage in the PBE-D3/cc-pVTZ calculations, resulting in an average bond length of 1.447 Å. This increase can be attributed to the incorporation of dispersion terms in the PBE-D3 method. These findings are consistent with a previous study [19], which reported bond lengths of 1.423, 1.531, 1.462 Å, and 1.369 Å for the optimized $D_{6d}$ structure. For the $O_h$ isomer, the carbon-carbon bond lengths were determined to be 1.3782 and 1.4957 Å at the B3LYP/cc-pVDZ level, in agreement with the MP2/6-31G calculations [19]. Regarding the angle measurements, the $D_{6d}$ isomer exhibits angles ranging from 106.7° to 120.0°, while the $O_h$ isomer displays angles within the range of 119.9° to 89.9°. The angles within the pentagons of both cage isomers deviate slightly from the ideal value of 108° for a regular pentagon. However, the angles within the hexagonal polygons closely approximate the expected value of 120° for a regular hexagon. For instance, at the PBE-D3 level, the angle A (1,2,10) of the $D_{6d}$ isomer measures 106.58°, which is 1.58° less than the regular pentagon angle, while the A(1,3,5) angle measures 109.86°, exceeding the regular pentagon angle by 1.86°. The angles within the quadrilaterals and hexagons of the $O_h$ isomer are nearly equal to the expected values of 120° and 90°, respectively. Furthermore, by examining the results in Tables 1 and 2, it can be observed that the empirically corrected dispersion terms (B3LYP-D3) have minimal impact on the B3LYP geometry.

**Table 3.** Bond lengths (r in Angstrom) and angles' measurements (A in degrees) at PBE-D3/cc-pVTZ, B3LYP/cc-pVTZ, B3LYP-D3/cc-pVTZ, MP2/cc-pVTZ and MP2/6-31G levels for the Ring $D_{12h}$.

| Isomer | $D_{12h}$ Ring | | | | | | |
|---|---|---|---|---|---|---|---|
| Basis set | cc-pVTZ | | | | 6-31G | HF/DZP [20] | MP2/DH(d) [21] |
| Methods | PBE-D3 | B3LYP | B3LYP-D3 | MP2 | MP2 | | |
| r(16-4) | 1.327 | 1.336 | 1.335 | Unstable | Unstable | 1.385 | 1.345 |
| r(4-15) | 1.248 | 1.228 | 1.227 | | | 1.197 | 1.266 |
| A(4-15-3) | 165.01 | 164.95 | 164.95 | - | - | - | 165.00 |
| A(3-14-2) | 164.99 | 165.06 | 165.06 | - | - | - | - |

**Table 4.** Bond lengths (r in Angstrom) and angles' measurements (A in degrees) at PBE-D3/cc-pVTZ, B3LYP/cc-pVTZ, B3LYP-D3/cc-pVTZ, MP2/cc-pVTZ and MP2/6-31G levels for the Sheet $D_{6h}$.

| Isomer | $D_{6h}$ Sheet | | | | | | |
|---|---|---|---|---|---|---|---|
| Basis set | cc-pVTZ | | | | 6-31G | HF/DZP [20] | B3LYP/ cc-pVDZ [19] |
| Methods | PBE-D3 | B3LYP | B3LYP-D3 | MP2 | MP2 | | |
| r(1-2) | 1.444 | 1.442 | 1.441 | 1.443 | 1.466 | 1.391 | 1.391 |
| r(6-12) | 1.496 | 1.483 | 1.482 | 1.485 | 1.495 | 1.456 | 1.488 |
| r(12-24) | 1.385 | 1.383 | 1.383 | 1.384 | 1.409 | 1.447 | 1.450 |
| r(17-24) | 1.241 | 1.228 | 1.228 | 1.254 | 1.275 | 1.210 | 1.240 |
| A(6-1-2) | 120.00 | 120.01 | 120.01 | 120.00 | 119.82 | 112.60 | - |
| A(1-2-3) | 119.99 | 119.99 | 119.99 | 120.00 | 119.82 | 127.40 | - |
| A(6-12-24) | 112.13 | 112.13 | 112.17 | 112.76 | 113.19 | 120.00 | - |
| A(10-16-23) | 127.86 | 127.86 | 127.82 | 127.23 | 126.69 | - | - |

Tables 3 and 4 present the bond lengths and angles for the $D_{12h}$ (Ring) and $D_{6h}$ (Sheet) isomers at the PBE-D3/cc-pVTZ, B3LYP/cc-pVTZ, B3LYP-D3/cc-pVTZ, MP2/cc-pVTZ and MP2/6-31G levels. The ring isomer $D_{12h}$ exhibits two types of bond lengths at the PBE-D3, B3LYP and B3LYP-D3 levels, while it is unstable at both the optimized MP2 levels, with either 6-31G or cc-pVTZ basis set. On the other hand, the sheet isomer $D_{6h}$ displays four different types of bonds across all four methods, where the calculations of the MP2 method were done for two different basis sets. The angles in the $D_{12h}$ isomer fall within the range of 164.9° to 165.0°, while the angles in the $D_{6h}$ isomer range from 112.28° to 127.71°. These angle values deviate slightly from those of a regular hexagon, which is consistent with previous studies [19,20,21]. Furthermore, when comparing the results of B3LYP-D3 and B3LYP for both $D_{12h}$ and $D_{6h}$ isomers, it can be observed that the dispersion correction only introduces minor changes. However, when we compare the optimized results of MP2/6-31G and MP2/cc-pVTZ, we find that the differences are small but significant, clearly distinguishing between the two types of calculations. It is worth noting that transitioning from the 6-31G basis set to the cc-pVTZ basis set has the effect of reducing the values of bond lengths and angles. Moreover, it is noteworthy that the results obtained with MP2/cc-pVTZ are very close to those of PBE and B3LYP-D3, with the exception of the value of r (17-24).

Table 5 provides the bond lengths and angle measurements for the $D_{2d}$ isomer at the PBE-D3/cc-pVTZ, B3LYP/cc-pVTZ, B3LYP-D3/cc-pVTZ and MP2/6-31G levels. The $D_{2d}$ isomer exhibits three types of bond lengths with an average length of 1.470 Å at the PBE-D3/cc-pVTZ level of computation, being slightly reduced to 1.465 Å at the B3LYP/cc-pVTZ level, whereas the B3LYP-D3/cc-pVTZ level yields the shortest average bond length of 1.462 Å, showing that the dispersion correction introduced by B3LYP-D3 causes only minor modifications. The angles in this isomer range from 88° to 150°. However, at the MP2 level, the $D_{2d}$ isomer is in a transitional state with a negative frequency of -313.33 cm$^{-1}$. While the B3LYP/cc-pVTZ method used in our study does not explicitly consider dispersion interactions, it is still considered a reliable functional for studying various molecular systems. Indeed, the B3LYP-D3 geometry is similar and nearly identical to that of B3LYP, which reduces the impact of the dispersion correction on the $D_{2d}$ structure. The negligible influence due to dispersion results in similar results of both B3LYP and B3LYP-D3 regarding energy and electrical properties. This finding suggests that the PBE-D3 method used in our analysis not only uses a reliable functional but also effectively incorporates dispersion forces and is specifically designed to accurately capture non-covalent interactions in molecular structures. Although there is limited experimental data available for the $C_{24}$ isomers, we recommend the use of the PBE-D3 method, which is particularly well-suited for investigating systems where dispersion and long-range interactions play a crucial role, such as the isomers under study. Therefore, the stability prediction of the $D_{2d}$ and $D_{12h}$ isomers by the PBE-D3 method, which takes into account dispersion interactions and non-covalent interactions, provides further support for its potential stability compared to the instability observed in the MP2 method.

**Table 5.** Bond lengths (r in Angstrom) and angles' measurements (A in degrees) at PBE-D3/cc-pVTZ, B3LYP/cc-pVTZ, B3LYP-D3/cc-pVTZ and MP2/6-31G levels for the $D_{2d}$ isomer (TS means "Transition State").

| Isomer | $D_{2d}$ Bracelet | | | | |
|---|---|---|---|---|---|
| Basis set | cc-pVTZ | | | | 6-31G |
| Method | PBE-D3 | B3LYP | B3LYP-D3 | MP2 | MP2 |
| r(4-13) | 1.482 | 1.483 | 1.478 | - | TS |
| r(13-6) | 1.393 | 1.379 | 1.380 | - | |
| r(13-5) | 1.525 | 1.517 | 1.527 | - | |
| A(4-13-6) | 150.01 | 150.03 | 150.02 | - | TS |
| A(6-23-8) | 149.95 | 149.88 | 149.92 | - | |
| A(24-6-13) | 91.67 | 91.97 | 91.85 | - | |

| A(13-5-24 ) | 88.33 | 88.02 | 88.14 | - | |

The average values of single, double, and triple bond lengths in the studied isomers are reported in Table 6. Upon examination of this table, it can be observed that for the $D_{6d}$, $D_{2d}$, and $D_{12h}$ isomers, the PBE-D3 method yields result that are nearly equivalent to those obtained with the B3LYP and B3LYP-D3 methods, while the MP2 method slightly overestimates the bond lengths compared to the previous three methods. In the case of the $D_{6h}$ isomer, both the PBE-D3 and B3LYP methods show almost identical results, while the MP2 method underestimates the bond lengths when compared to them. For the two isomers, $O_h$ and $D_{6h}$, both MP2/6-31G and MP2/cc-pVTZ methods yield average bond lengths of similar orders of magnitude but with significantly different values. This clearly demonstrates the impact of expanding the basis set**.**

**Table 6.** The average values of the simple and multiple bond lengths at PBE-D3/cc-pVTZ, B3LYP/cc-pVTZ, B3LYP-D3/cc-pVTZ, MP2/cc-pVTZ and MP2/6-31G levels for $C_{24}$ isomers.

| Basis set | Method | Distance | $D_{6d}$ | $O_h$ | $D_{12h}$ | $D_{6h}$ | $D_{2d}$ |
|---|---|---|---|---|---|---|---|
| cc-pVTZ | PBE-D3 | $\bar{r}_{c-c}$ | 1.466 | 1.436 | 1.247 | 1.440 | 1.459 |
| | | $\bar{r}_{c=c}$ | 1.428 | 1.494 | - | 1.363 | 1.482 |
| | | $\bar{r}_{c\equiv c}$ | - | - | 1.327 | 1.241 | - |
| | B3LYP | $\bar{r}_{c-c}$ | 1.458 | 1.458 | 1.228 | 1.434 | 1.448 |
| | | $\bar{r}_{c=c}$ | 1.434 | 1.489 | - | 1.359 | 1.483 |
| | | $\bar{r}_{c\equiv c}$ | - | - | 1.336 | 1.228 | - |
| | B3LYP-D3 | $\bar{r}_{c-c}$ | 1.462 | 1.431 | 1.227 | 1.433 | 1.453 |
| | | $\bar{r}_{c=c}$ | 1.436 | 1.494 | - | 1.358 | 1.478 |
| | | $\bar{r}_{c\equiv c}$ | - | | 1.335 | 1.228 | - |
| | MP2 | $\bar{r}_{c-c}$ | - | 1.435 | Unstable | 1.436 | - |
| | | $\bar{r}_{c=c}$ | - | 1.483 | | 1.366 | - |
| | | $\bar{r}_{c\equiv c}$ | - | - | | 1.254 | - |
| 6-31G | MP2 | $\bar{r}_{c-c}$ | 1.489 | 1.402 | Unstable | 1.357 | TS |
| | | $\bar{r}_{c=c}$ | 1.465 | 1.520 | | 1.389 | |
| | | $\bar{r}_{c\equiv c}$ | - | - | | 1.275 | |

### 3.2. Stability, Gap, Fermi and Atomization energies

### Relative Stability

All four methods agree that the $D_{6h}$ (Sheet) isomer has the lowest energy, ranking it in fact as the most stable. The Fig. 2 illustrates the relative energy histogram which reflects the chemical stability of the different isomers compared to the most stable one $D_{6h}$, as determined by the four methods. According to the calculations using the PBE-D3/cc-pVTZ, B3LYP/cc-pVTZ and B3LYP-D3/cc-pVTZ methods, the newly proposed $D_{2d}$ structure is predicted to be stable. There is a significant difference in the relative stability of the $D_{2d}$ structure between the PBE-D3/cc-pVTZ and B3LYP/cc-pVTZ methods, with a gap of approximately 13%. This difference can be solely attributed to the choice of the DFT functional, as the basis set remains the same for both methods. Contrary to the observations regarding the geometry, when it comes to the relative stability predicted by B3LYP and B3LYP-D3 using the same basis set, both methods exhibit sometimes significant discrepancies depending on the isomer. For the $D_{2d}$ and $O_h$ isomers, the relative energy gap between the two methods is 0.03% and 3.3%, respectively. However, for $D_{12h}$ and $D_{6d}$, this difference becomes much larger, attaining 64.9% and 70.5%, respectively. From this, we conclude that the dispersion effects, unlike the geometry case where they play a minor role, are crucial factors in determining the relative energies. However, it is worth noting that the $D_{2d}$ structure is ranked last among the other structures. This observation could potentially be explained by its higher rigidity and stronger geometric constraints compared to the other geometric forms. However, most likely due to the use of the non-extended 6-31G basis set, the MP2 method predicts the $D_{2d}$ structure to be a transition state.

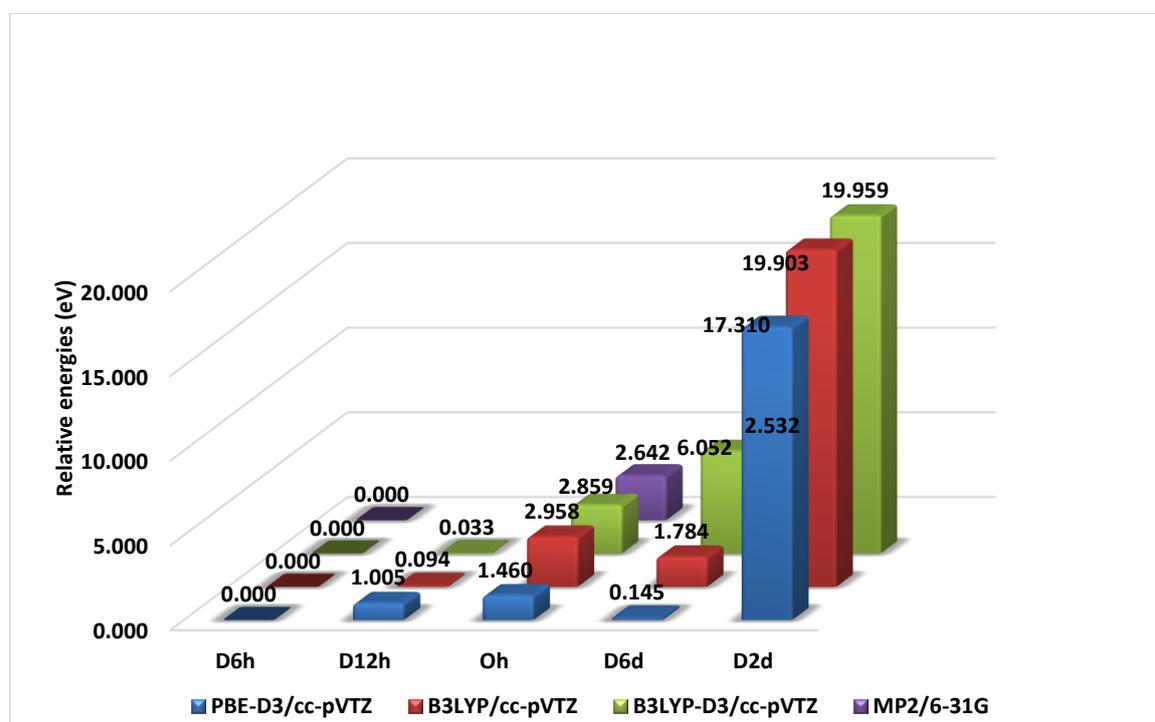

**Fig. 2**. Relative energies (eV) of the $C_{24}$ isomers at PBE-D3/cc-pVTZ, B3LYP/cc-pVTZ, B3LYP-D3/cc-pVTZ and MP2/6-31G levels. *

---

* A violet cube, corresponding to MP/6-31G method, of height 2.532 exists, but is not visible, for the isomer $D_{6d}$.

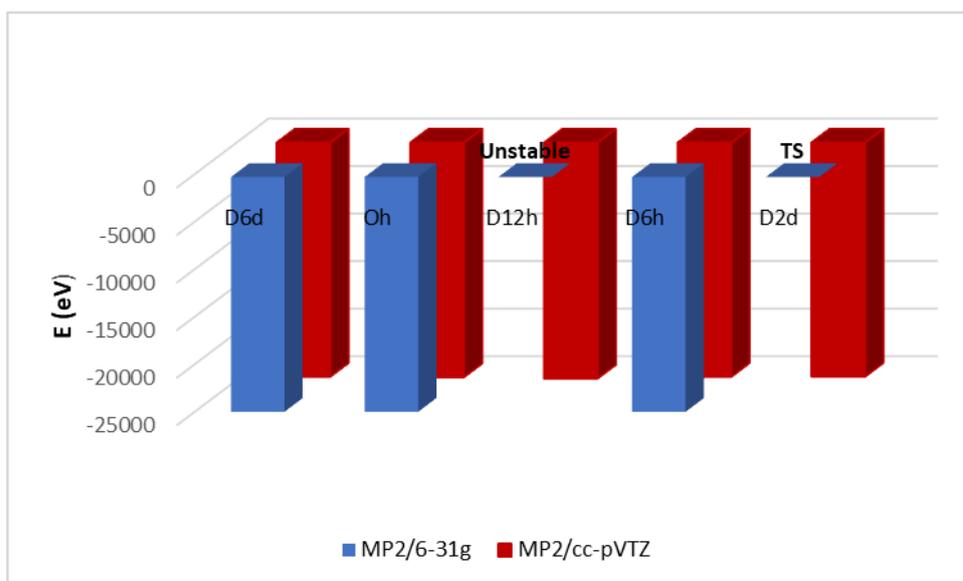

**Fig. 3. Comparison in MP2 method between** Optimized Energy calculations using the 6-31G for all five isomers, predicting $D_{2d}$ to be TS and $D_{12h}$ to be unstable, and those using the cc-pVTZ for three isomers ($O_h$, $D_{12h}$ and $D_{6h}$). Also shown are the single point calculation of MP2/cc-pVTZ for the two isomers ($D_{2d}$ and $D_{6d}$).

We illustrate in Fig. 3 a comparison, in the MP2 method, between the optimized energies using the basis set 6-31G for all isomers, where it predicts the isomer $D_{2d}$ to be a transition state (TS) and $D_{12h}$ to be unstable, and those using the basis set cc-pVTZ for the isomers ($O_h$, $D_{12h}$, $D_{6h}$), whereas we list also the single point calculation using the cc-pVTZ basis set in the two isomers ($D_{2d}$ and $D_{6d}$) where convergence for optimization was not guaranteed. Actually, the figure reveals that while the MP2/6-31G optimization predicts an unstable $D_{12h}$ isomer and a transition state for $D_{2d}$, the energy differences between the two methods are minimal for the other isomers. This implies and explains that significant energy changes can only occur after numerous iterations. Therefore, caution must be exercised when interpreting the results from a single MP2/cc-pVTZ calculation.

## Atomization Energies

Table 7 and Figures 4 and 5 present the atomization energies (AE), Fermi energies (FE), and gap energies (GE) of the $C_{24}$ isomers obtained from calculations performed at four different levels of theory: (PBE-D3, B3LYP, B3LYP-D3)/cc-pVTZ, and MP2/6-31G for the five isomers, as well as MP2/cc-pVTZ for the two isomers $O_h$ and $D_{6h}$. The atomization energies were determined using specific formulas:

$$AE = \frac{24 \cdot E_{Carbon} - E_{Molecule}}{24}$$

**Table 7.** Atomization energy (in eV/atom) of $C_{24}$ isomers at different calculations levels.

| AE (eV/atom) | | | | | | |
|---|---|---|---|---|---|---|
| Basis set | Method | $D_{6d}$ | $O_h$ | $D_{12h}$ | $D_{6h}$ | $D_{2d}$ |
| cc-pVTZ | PBE-D3 | 8.729 | 8.670 | 8.690 | 8.340 | 8.010 |
| | B3LYP | 8.005 | 7.956 | 8.076 | 8.080 | 7.246 |
| | B3LYP -D3 | 7.825 | 7.958 | 8.076 | 8.077 | 7.246 |
| 6-31G | MP2 | 6.200 | 6.196 | Unstable | 6.309 | TS |
| cc-pVTZ | | - | 6.534 | - | 6.636 | - |
| B3LYP/6-31+G* [25] | | 9.03 | - | - | - | - |

It is important to note that the MP2 method consistently yields lower atomization energy (AE) values compared to the PBE and B3LYP methods for each isomer. The average value of AE obtained from DFT-type methods (PBE-D3, B3LYP and B3LYP-D3) is 8.06 eV/atom, while the MP2 method with cc-pVTZ (6-31G) basis set provides an average value of 6,59 (6.24) eV/atom, which is approximately 23.6% lower than the DFT methods. Specifically, for the $D_{6d}$ isomer, the average value of the AE using DFT methods (PBE-D3, B3LYP and B3LYP-D3) is 8.19 eV/atom, which is relatively close to the reference [25] calculated using the same method but through a different basis set, with a deviation of approximately 9.3%. Furthermore, it is worth noting that the PBE-D3 method consistently yields atomization energies AE that are higher than those obtained from the B3LYP and B3LYP-D3 methods by approximately 7.5%. Regarding the new $D_{2d}$ structure, it is observed that its atomization energy AE in the PBE-D3 method deviates from the closest isomer, $D_{6h}$, by only 0.33 eV per atom. However, both the B3LYP and B3LYP-D3 methods distinctly separate the $D_{2d}$ isomer from the $D_{6d}$ isomer, exhibiting a noticeable deviation of 0.83 eV per atom for both methods.

## Gap Energies

The gap energy (GE), is defined as the minimum amount of energy needed for an electron to move from the valence band to the conduction band, indicating whether the materials are conductors, semiconductors, or insulators. It is known [40], that a material is an insulator when the GE is larger than 3 eV, whereas it is a semi-conductor when the GE is moderately large between 0.1 to 3 eV, which is lower than in an insulator and is of the order of (~1eV). For a conductor, conduction bands and valence bands are not separated and the GE is therefore vanishing.

Fig. 4 illustrates that the GE provided by the MP2 method (using cc-pVTZ and 6-31G basis set) is significantly larger than that of the PBE-D3, B3LYP and B3LYP-D3 methods. Furthermore, it is worth noting that both B3LYP and B3LYP-D3 methods produce almost identical GEs, but these gaps are 56.65% larger than those obtained from the PBE-D3 method. Specifically, the average DFT (PBE, B3LYP and B3LYP-D3) GE is 1.89 eV, whereas it is 7.63 (7.97) eV in the MP2/6-31G (/cc-pVTZ), representing an increase of approximately 75,7%. This substantial difference can be attributed to the use of the non-extended basis set in the MP2 method, which leads to that all $C_{24}$ isomers behave as insulators. A similar trend is observed when comparing the two DFT methods, in that the average GE resulting from B3LYP and B3LYP-D3 methods is 2.33 eV, while the PBE-D3 method yields a value of 1.01 eV for the GE, representing an increase of 56.65%. The fact that the GE values of the B3LYP and the B3LYP-D3 methods are nearly identical suggests that the functional nature is likely the determining factor in the discrepancy between (B3LYP + B3LYP-D3) on one side and PBE-D3 on the other side.

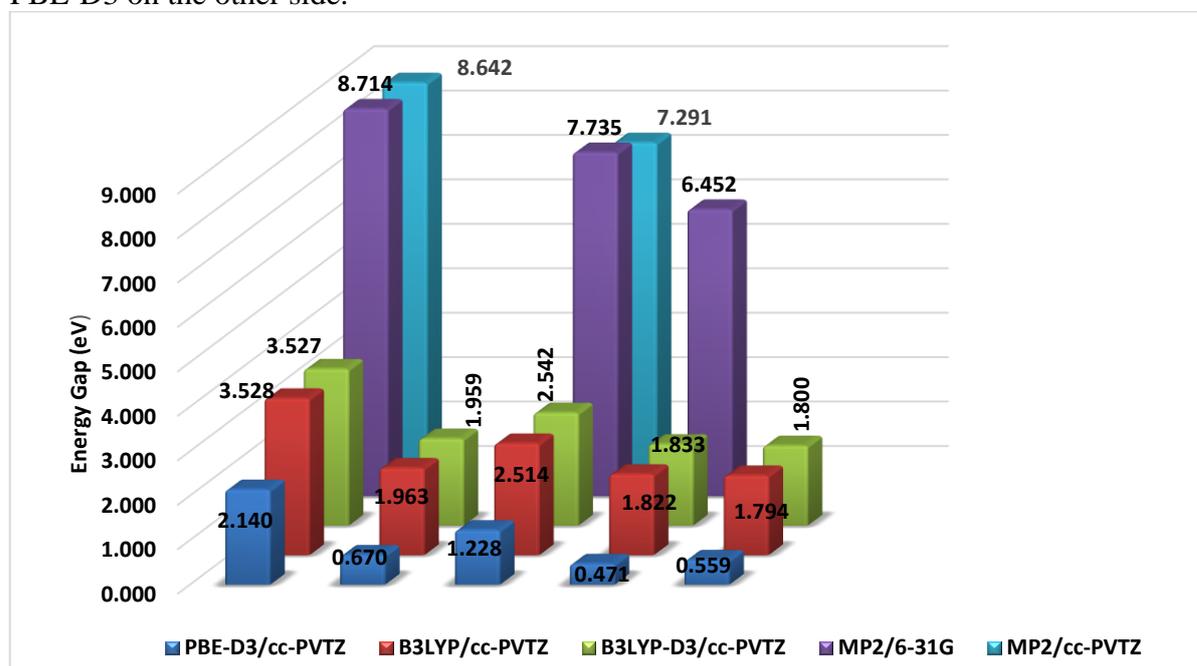

**Fig. 4**. The $E_{Gap}$ energy (eV) of $C_{24}$ isomers ($D_{6h}$, $D_{12h}$, $O_h$, $D_{6d}$ and $D_{2d}$, from left to right) at the PBE-D3, B3LYP, B3LYP-D3, MP2 (using both cc-pVTZ and 6-31G) levels.

The B3LYP-D3 method, as well as the B3LYP, notably overestimates the values of the GE compared to the lower values provided by PBE-D3. The average GE in the (B3LYP+B3LYP-D3) methods is approximately 2.32 eV, while it is 1.01 eV in the PBE-D3 method. Therefore, the (B3LYP+B3LYP-D3) methods yields GE values approximately 56.5% greater than those of PBE-D3. Both the PBE-D3 and (B3LYP+B3LYP-D3) methods agree that the $D_{6h}$ isomer has the highest GE, despite having the lowest relative energy. The (B3LYP+B3LYP-D3) method establishes a correlation between the GE and the relative energy for the $D_{2d}$ isomer, with the smallest GE in the (B3LYP+B3LYP-D3) method. However, this correlation is not maintained in the PBE method. We note that in all the methods used, the sheet isomer ($D_{6h}$) has the largest GE values compared to other isomers with the least electrical conductivity. According to the (B3LYP+B3LYP-D3) methods, the $D_{2d}$ isomer has the smallest GE, while the PBE-D3 method predicts it to be $D_{6d}$. With the exception of the $D_{6h}$ isomer which is to be considered a dielectric material with a GE of 3.53 eV and 2,14 eV in B3LYP and PBE respectively, the PBE and (B3LYP+B3LYP-D3) methods do not agree on the electrical conductivity of the C24 isomers. Specifically, according to the PBE-D3 method, the $D_{12h}$, $D_{6d}$, $D_{2d}$, and $O_h$ isomers are predicted to be semiconductors, while the (B3LYP+B3LYP-D3) methods classifies them as weak insulators. The MP2 method predicts a significantly clear insulating character for the $D_{6h}$, $O_h$, and $D_{6d}$ isomers, making the $D_{6h}$ isomer to be the most insulating among the five studied isomers. In comparison with other studies, we find an agreement when comparing with the previous theoretical calculations at B3LYP/6-31G (d) [27], giving

GE values of 1.825, 2.522, 1.881, and 3.425 eV for the $D_{6d}$, $O_h$, $D_{12h}$, and $D_{6h}$ isomers respectively. On the other hand, when comparing our calculations at MP2 methods with previous theoretical calculations at the MP2/DZP level [20], where GE values of 7.346, 7.619, 8.435, and 9.251 eV were measured for the same isomers, respectively, we find also a good agreement. On the other hand, the results of the works [27], showed that the $C_{24}$ semiconducting fullerene isomers, except for the $D_{6h}$ isomer, were also considered dielectric materials with a GE of 3.425 eV.

## Fermi Energies

In Figure 5, we present the Fermi energy (FE) values of the five isomers calculated using the formula:

$$E_F = (E_{LUMO} + E_{HOMO})/2$$

using the four methods (PBE-D3, B3LYP, B3LYP-D3)/cc-pTVZ and MP2/6-31G, as well as those of the $O_h$ and $D_{6h}$ isomers using MP2/cc-pTVZ.

These FE values are shown according to the energy stability of the B3LYP/cc-pVTZ method from left to right ($D_{6h}$, $D_{12h}$, $O_h$, $D_{6d}$ and $D_{2d}$ respectively).

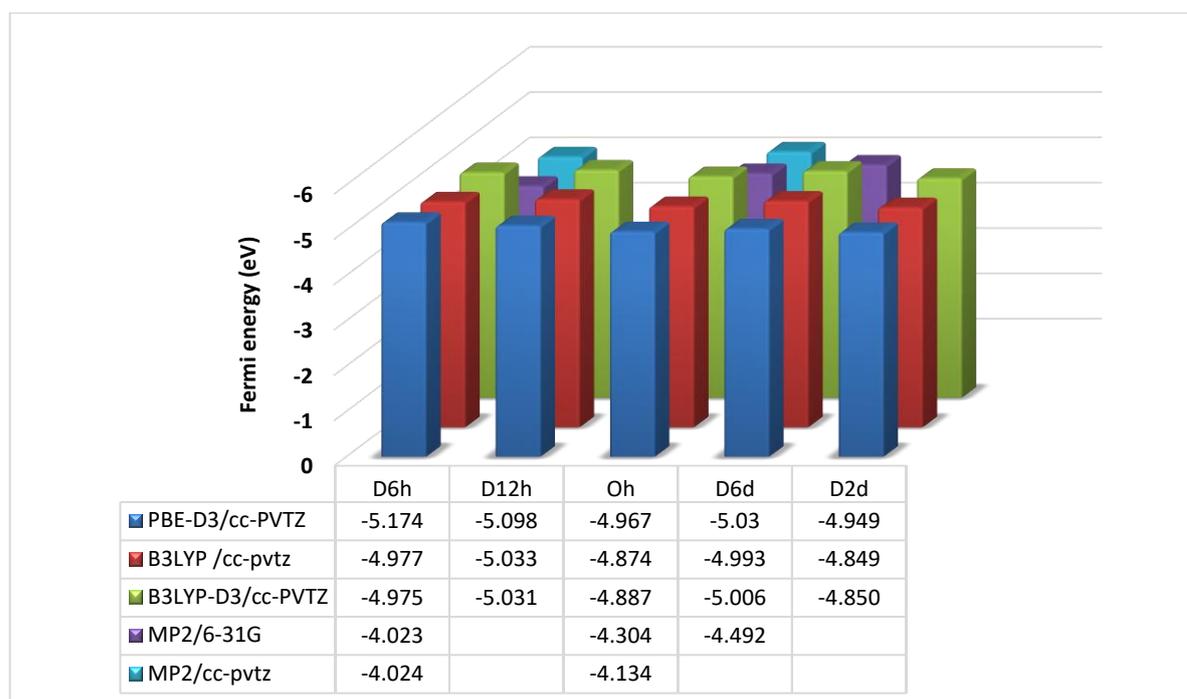

|  | D6h | D12h | Oh | D6d | D2d |
|---|---|---|---|---|---|
| PBE-D3/cc-PVTZ | -5.174 | -5.098 | -4.967 | -5.03 | -4.949 |
| B3LYP /cc-pvtz | -4.977 | -5.033 | -4.874 | -4.993 | -4.849 |
| B3LYP-D3/cc-PVTZ | -4.975 | -5.031 | -4.887 | -5.006 | -4.850 |
| MP2/6-31G | -4.023 |  | -4.304 | -4.492 |  |
| MP2/cc-pvtz | -4.024 |  | -4.134 |  |  |

**Fig. 5.** Fermi energies (in eV) of the isomers of $C_{24}$ at PBE-D3, B3LYP, B3LYP-D3, MP2 (using cc-pVTZ ) and MP2/6-31G levels.

Fig. 5 illustrates that the values of the FE for the five $C_{24}$ isomers calculated using the four methods are relatively comparable, exhibiting a contrasting behavior compared to the energy gap. From the figure 5, it is noticeable that the MP2 method consistently underestimates the FE values of the various stable isomers compared to the PBE-D3,

B3LYP and B3LYP-D3 methods. The average FE value predicted by the DFT methods (PBE-D3, B3LYP and B3LYP-D3) is 4,98 eV, whereas the MP2 method (using the two basis sets) provides a value of 4.20 eV, representing an increase of 15,7%. In contrast, the three DFT methods provide very similar values. The average FE value in the (B3LYP and B3LYP-D3) methods is 4.95, while in the PBE-D3 method it is 5.04, resulting in a 1.79% increase for PBE compared to (B3LYP and B3LYP-D3) methods. Additionally, both the PBE-D3 and (B3LYP and B3LYP-D3) methods agree that the $D_{2d}$ isomer has the highest FE value, but they do not agree on ranking the isomer with the highest FE. The PBE-D3 method ranks the $D_{6h}$ isomer as having the lowest FE, while (B3LYP and B3LYP-D3) assign this position to the $D_{12h}$ isomer. The ranking of the isomer with the highest and lowest FE values by the PBE and (B3LYP and B3LYP-D3) methods is consistent with the ranking of the atomization energy discussed earlier. Moreover, we observe that the $D_{6h}$ isomer has the highest FE at the MP2 level unlike the PBE-D3 and (B3LYP and B3LYP-D3) methods where it is the $D_{2d}$ isomer which meets this property.

### 3.3. Electronic Properties of $C_{24}$ Isomers: Ionization Potential, Electronic Affinity, Hardness, and Electronegativity

The computed values of the ionization potential (IP), electronic affinity (EA), hardness (η), and electronegativity (χ) of all isomers are given in Tables (8 and 9). For the MP2 method, these values can be calculated from the HOMO and LUMO orbital energies using the following approximate expression:

$$IP = -E_{Homo}, \quad EA = -E_{Lumo}$$

$$\text{Electronegativity } (\chi) = \frac{(IP+EA)}{2}$$

$$\text{Hardness } (\eta) = \frac{(IP-EA)}{2}$$

The IP and EA values are determined using the PBE-D3 and (B3LYP and B3LYP-D3) methods with the cc-pVTZ basis set, employing the respective expressions:

$$IP = E(\text{optimized cation}) - E(\text{optimized neutral})$$
$$EA = E(\text{optimized neutral}) - E(\text{optimized anion})$$

**Table 8.** Ionization potential (IP) and electronic affinity (EA), in eV, of the $C_{24}$ isomers at (PBE-D3, B3LYP, B3LYP -D3, MP2)/cc-pVTZ and MP2/6-31G levels.

| | | IP (eV) | | | | |
|---|---|---|---|---|---|---|
| Basis set | Method | $D_{6d}$ | $O_h$ | $D_{12h}$ | $D_{6h}$ | $D_{2d}$ |
| cc-pVTZ | PBE-D3 | 7.219 | 7.571 | 6.973 | 8.096 | 7.190 |
| | B3LYP | 6.426 | 7.719 | 7.222 | 8.159 | 7.093 |
| | B3LYP -D3 | 3.178 | 7.647 | 7.202 | 8.124 | 7.130 |
| 6-31G | MP2 | 7.718 | 8.172 | Unstable | 8.380 | - |
| cc-pVTZ | | - | 7.780 | - | 8.345 | - |
| 6-31G(d) | B3LYP [26] | 7.36 | 7.54 | - | 8.00 | - |
| | B3LYP [41] | 7.47 | - | - | - | - |
| EA (eV) | | | | | | |
| cc-pVTZ | PBE-D3 | 2.845 | 2.145 | 3.289 | 2.358 | 2.758 |
| | B3LYP | 2.474 | 2.110 | 2.936 | 1.911 | 2.609 |
| | B3LYP -D3 | 6.813 | 2.058 | 2.869 | 1.939 | 2.584 |
| 6-31G | MP2 | 1.265 | 0.436 | - | -0.333 | - |
| cc-pVTZ | | - | 0.489 | - | -0.296 | - |
| 6-31G(d) | B3LYP [41] | 2.98 | - | - | - | - |
| Exp. [42] | | 2.90 | - | - | - | - |

**Table 9.** Hardness (η) and electronegativity(χ), in eV, of the isomers of $C_{24}$ at (PBE-D3, B3LYP, B3LYP-D3, MP2)/cc-pVTZ and MP2/6-31G levels.

| η (eV) | | | | | | |
|---|---|---|---|---|---|---|
| Basis set | Methods | $D_{6d}$ | $O_h$ | $D_{12h}$ | $D_{6h}$ | D2d |
| cc-pVTZ | PBE-D3 | 2.187 | 2.713 | 1.842 | 2.869 | 2.216 |
| | B3LYP | 1.976 | 2.804 | 2.143 | 3.124 | 2.242 |
| | B3LYP -D3 | 1.817 | 2.794 | 2.166 | 3.092 | 2.273 |
| cc-pVTZ | MP2 | - | 3.645 | - | 4.320 | - |
| 6-31G | | 3.226 | 3.868 | Unstable | 4.356 | TS |
| 6-31G(d) | B3LYP[41] | 0.89 | - | - | - | - |
| χ (eV) | | | | | | |
| cc-pVTZ | PBE-D3 | 5.032 | 4.858 | 5.131 | 5.227 | 4.974 |
| | B3LYP | 4.450 | 4.914 | 5.079 | 5.035 | 4.851 |
| | B3LYP -D3 | 4.995 | 4.852 | 5.035 | 5.031 | 4.857 |
| cc-pVTZ | MP2 | - | 4.134 | Unstable | 4.024 | - |
| 6-31G | | 4.491 | 4.304 | - | 4.023 | - |
| 6-31G(d) | B3LYP[41] | 5.19 | - | - | - | - |

As shown in Tables 8 and 9, the IP, EA, η and χ values for all the five isomers were calculated using the DFT (PBE-D3, B3LYP, B3LYP-D3)/cc-pVTZ and the MP2/6-31G. In addition, the same values for the $O_h$ and $D_{6h}$ isomers using the MP2/cc-pVTZ level are stated.

The results indicate that, excluding the $D_{6d}$ isomer at the B3LYP-D3 level, the IP values for the five isomers are similar among all four methods, consistent with a previous study [26] (and also [41]) which, using the B3LYP/6-31G(d) method, reported an average IP value of 7.42 eV, slightly lower than our average of 7.51 eV obtained through the DFT calculations. Furthermore, upon examining Table 8, several observations can be noted. For the isomers $O_h$, $D_{12h}$, $D_{6h}$, and $D_{2d}$, both the B3LYP and B3LYP-D3 methods yield almost identical Ips, which implies, for these isomers, that the dispersion correction has only a minor impact. In other words, the inclusion of the dispersion correction does not significantly alter the IP values obtained using the B3LYP method for these specific isomers.

Moreover, for all the isomers, the MP2 method (using 6-31G and cc-pVTZ basis sets) overestimates the IP values compared to those obtained from the DFT methods (B3LYP, B3LYP-D3, and PBE-D3). In simpler terms, the MP2 method, with its respective basis sets, tends to provide higher IP values for all isomers when compared to those obtained from the DFT methods (B3LYP, B3LYP-D3 and PBE-D3). Except for $D_{6d}$, the MP2 (6-31G and cc-pVTZ) method gives an average IP value of 8.17 eV, while DFT methods (B3LYP, B3LYP-D3, and PBE-D3) yield a lower average of 7.51 eV, showing an 8.1% decrease. For $D_{6d}$ isomer, the MP2 (6-31G and cc-pVTZ) method gives an average IP of 7.39 eV, while the PBE-D3 and B3LYP methods yield 6.82 eV and the B3LYP-D3 gives 3.18 eV. The dispersion correction appears to have a significant impact on the IP value as it reduces the IP by 53.4%. The effect of expanding the basis set from 6-31G to cc-pVTZ on the MP2 calculations of the IP is minor for the $D_{6h}$ isomer, whereas a substantial decrease is observed for the $O_h$ isomer.

Putting aside the $D_{6h}$, the $C_{24}$ isomers exhibit positive EA values, indicating their ability to form stable anions. Among these isomers, $D_{12h}$ shows the highest EA. Comparing the PBE-D3/cc-pVTZ and B3LYP/cc-pVTZ levels, the EA of $D_{6d}$ is determined to be 2.845 eV and 2.474eV, respectively, which closely matches an experimental study reporting 2.90 eV [42], while a different theoretical approach using B3LYP/6-31G(d) yielded a slightly higher value of 2.98 eV [41]. Table 8 indicates that for the $D_{6d}$ isomer, the EA provided by the PBE-D3 method is slightly higher than that provided by B3LYP. The average EA value for these two methods is 2.66 eV. However, the B3LYP-D3 method significantly overestimates the EA, giving a value of 6.81 eV, which is represents a 60.9% increase compared to (PBE and B3LYP). Additionally, comparing the values provided by B3LYP and B3LYP-D3 suggests that dispersion terms lead to a 60.9% increase in the EA value, while for other isomers, their contribution is minor. Regarding the MP2 findings, it is evident that the MP2/6-31G method significantly underestimates the EA in comparison to the three DFT methods. This discrepancy is likely attributed to the constrained nature of the basis set employed, which might not fully capture the intricate electronic structure of the system. The results from Table 8 indicate that the PBE-D3 method slightly overestimates the EA values for the $O_h$, $D_{12h}$, $D_{6h}$, and $D_{2d}$ isomers. Furthermore, the average EA obtained by the three DFT methods (PBE-D3, B3LYP, and B3LYP-D3) is 2.46 eV. Notably, the two MP2 methods (utilizing 6-31G and cc-pVTZ) yield very similar EA values. Specifically, the average EA for the $O_h$ isomer is 0.46 eV, and for $D_{6h}$, it is -0.31 eV.

Hardness reflects a molecule's resistance to changes in electron distribution, with "hard" molecules, like $D_{6d}$ with a large HOMO-LUMO gap, being less reactive. Conversely, softness suggests higher reactivity. Moreover, hardness indicates a molecule's ability to resist the deformation of its electron cloud during chemical processes.

According to Table 9, the three DFT methods (PBE-D3, B3LYP, and B3LYP-D3) yield hardness ($\eta$) and electronegativity ($\chi$) values of comparable magnitude. The average values for each isomer and for the isomer series are as follows: (hardness; electronegativity) - (1.99; 4.83) for $D_{6d}$, (2.77; 4.61) for $O_h$, (2.05; 5.08) for $D_{12h}$, (3.03; 5.10) for $D_{6h}$, (2.24; 4.89) for $D_{2d}$, and (2.42; 4.90) for the isomer series.

On the other hand, the $D_{6h}$ isomer demonstrates the highest hardness and electronegativity ($\chi$) values at the PBE-D3, B3LYP, B3LYP-D3 and MP2 (utilizing cc-pVTZ and 6-31G) levels. Furthermore, the use of the MP2 method with cc-pVTZ and 6-31G leads to a significant overestimation of $\eta$. For the $O_h$ and $D_{6h}$ isomers, the MP2 method predicts $\eta$ values of 3.65 eV and 4.32 eV, respectively, while the average $\eta$ obtained from the DFT methods (PBE-D3, B3LYP, and B3LYP-D3) is 2.77 eV and 3.03 eV. This corresponds to an increase of 24% and 29.9%, respectively. Similar trends are observed when applying the MP2/6-31G method. In particular, for the $D_{6d}$ isomer, the average $\eta$ and $\chi$ obtained from the three DFT methods are 1.99 eV and 4.83 eV, respectively. However, other studies [41], which employed the B3LYP/6-31G(d) method, report significantly different values of 0.89 eV and 5.19 eV, resulting in discrepancies of 55.28% and 6.93%, respectively. In contrast, the MP2/6-31G method yields values of 3.23 eV and 4.49 eV, leading to deviations of 72.45% and 13.49%, respectively. These findings emphasize that despite having the same number of carbon atoms (24), the isomers differ in their molecular geometry, which accounts for the variation in their physical properties. Therefore, molecular geometry plays a vital role in determining the physicochemical properties of substances.

### 3.4. Electric Dipole Moment and Polarizabilities

Tables (10 and 11) present the results of electric dipole moment ($\mu$) and static electric polarizability $<\alpha>$ for the $C_{24}$ isomers calculated at the PBE-D3, B3LYP, B3LYP-D3 and MP2, using the cc-pVTZ basis set and 6-31G basis set respectively.

### Electric Dipole Moment

The electric dipole moment is a vector quantity that is significantly affected by the molecular symmetry and flexibility. It characterizes both the magnitude and direction of the electric charge separation within a molecule, and it is equal to the vector sum of the electric dipole moments of its individual bonds. When a molecule exhibits perfect symmetry in its structure and charge distribution, the individual bond moments cancel each other out, causing the overall dipole moment to vanish. However, as the molecule moves away from symmetry, the dipole moment increases, reflecting the growing charge separation within the molecule.

In Table (10) we show the dipole moment values calculated at the PBE-D3, B3LYP, B3LYP-D3, MP2 (utilizing cc-pVTZ) and the MP2/6-31G levels.

**Table 10.** Electric dipole moment (Debye) of the five $C_{24}$ isomers at PBE-D3, B3LYP, B3LYP -D3, MP2 (using cc-pVTZ) and MP2/6-31G levels.

| Dipole Moment ($\mu$)×$10^{-4}$ | | | | | | |
|---|---|---|---|---|---|---|
| Basis set | Method | $D_{6d}$ | $O_h$ | $D_{12h}$ | $D_{6h}$ | $D_{2d}$ |
| cc-pVTZ | PBE-D3 | 366.4 | 0. 1 | 0. 3 | 0. 6 | 0. 1 |
| | B3LYP | 11.5 | 0. 1 | 6.2 | 0. 2 | 0. 1 |
| | B3LYP -D3 | 7.6 | 0. 1 | 6.8 | 0. 2 | 0. 1 |
| 6-31G | MP2 | 12.7 | 0.5 | Unstable | 0.0 | TS |
| cc-pVTZ | | | - | 0.0 | - | 0.0 | - |
| 6–311 + + G(d,p) | B3LYP) [43] | 0 | - | - | - | - |

According to Table 10, the dipole moment μ values for the $D_{6d}$ isomer vary significantly among the four methods used in this study and the one utilized in an earlier work [43]. The differences in value range up to 97.9% when compared to our lowest value and to 100% compared to the B3LYP/6-311++G(d,p) result [43]. PBE-D3 seems to poorly represent μ compared to B3LYP and B3LYP-D3. However, the inclusion of dispersion terms in B3LYP-D3 lowers μ by 33.9% compared to B3LYP. The MP2/6-31G method provides a μ value that is 9.4% (40.2%) higher than that of B3LYP (B3LYP-D3). Considering the basis set extension used in [43], the value of 7.6 Debye (D) appears to be the most plausible among our results. Concerning the $O_h$ isomer, the numerical complexity is reduced because all three DFT methods unexpectedly give the same μ value of 0.1 D, even though they use different functionals and account for dispersion in different ways. However, when employing the MP2/6-31G method, the dipole moment μ increases by 80% compared to the results obtained from the DFT methods.

In the case of the $D_{12h}$ isomer, when incorporating dispersion terms in B3LYP, the dipole moment increases by 8.8% compared to the result obtained solely with the B3LYP method. The average dipole moment obtained from both B3LYP and B3LYP-D3 is 6.5 D, showing that it is 95.4% higher than the dipole moment obtained from the PBE-D3 method. In the case of $D_{6h}$, the situation is comparable in nature to $D_{12h}$. Adding dispersion terms to the B3LYP method does not have a notable impact on the results. The average dipole moment obtained from both B3LYP and B3LYP-D3 is 0.2 D due to its high symmetry, which is 66.7% lower than the dipole moment value obtained from the PBE method.

Lastly, we discuss the new $D_{2d}$ bracelet isomer, where all three DFT methods yield identical μ values. The influence of functionals nature and dispersion terms seems to be negligible in this scenario. Furthermore, despite the diverse range of geometries observed among the isomers, the dipole moment remains constant during the transition from one isomer to another, which appears to be an unreasonable outcome. The conclusion drawn from this discussion regarding the dipole moment variation along the $C_{24}$ isomer series reveals that despite the dipole moment being a first-order property, it is not straightforward to accurately replicate it. The selection of the functional and the approach employed to incorporate long-range interactions are determining factors in achieving accurate results. The three DFT methods indicate that $D_{2d}$ and $O_h$ isomers have equal and lowest dipole moments, whereas the B3LYP/6-311++G(d,p) [43] method suggests that $D_{6d}$ occupies this position. It is evident that the three DFT methods (using cc-pVTZ) and MP2/6-31G concur that $D_{6d}$ possesses the largest dipole moment in the isomer series, contrary to the previous study [43], which predicted $D_{6d}$ to have the smallest dipole moment among all the isomers.

## Atomic and Molecular static Polarizabilities

Table (11) presents the isotropic polarizability (<α>) of the stable $C_{24}$ isomers calculated at the DFT (PBE-D3, B3LYP and B3LYP-D3) and MP2 levels using the cc-pVTZ basis set and 6-31G basis set respectively. The isotropic static electric polarizability <α> is calculated as the mean value of those for the three molecule axes as expressed by:

$$<\alpha> = \frac{\alpha_{xx} + \alpha_{yy} + \alpha_{zz}}{3}$$

**Table 11.** Comparison of the static electric polarizability values (evaluated in atomic units) for the $C_{24}$ isomers at the (PBE-D3, B3LYP, B3LYP-D3)/cc-pVTZ and MP2/6-31G levels.

| | | Molecular Polarizability (a.u) | | | | | Atomic Polarizability (a.u) | |
|---|---|---|---|---|---|---|---|---|
| Basis set | Method | $D_{6d}$ | $O_h$ | $D_{12h}$ | $D_{6h}$ | $D_{2d}$ | $\alpha_C$ | $24*\alpha_C$ |
| cc-pVTZ | PBE-D3 | 219.02 | 202.89 | 548.08 | 269.39 | 265.20 | 8.05 | 193.37 |
| | B3LYP | 201.23 | 198.85 | 498.59 | 259.19 | 259.40 | 7.91 | 189.92 |
| | B3LYP -D3 | 201.92 | 199.40 | 498.26 | 259.01 | 260.52 | 7.91 | 189.92 |
| 6-31G | MP2 | 189.33 | 172.57 | Unstable | 223.27 | TS | 5.07 | 121.82 |
| 6-31+G(d) | B3LYP [44] | 215 | - | - | - | - | - | - |
| 6-311+G(d) | PBEPBE [45] | 187.13 | - | - | - | - | - | - |
| 6–311 + + G(d,p) | B3LYP [43] | 216.19 | - | - | - | - | - | - |
| Atomic Polarizability (a.u) (Previous work) | | | | | | | | |
| Comments | | $\alpha_C$ | | Ref. | | | | |
| NR, CASPT2, ML res.<br>NR, CCSD(T), ML res.<br>R, Dirac+Gaunt,CCSD(T)<br>recommended | | 11.39<br>11.67±0.07<br>11.26±0.20<br>11.3±0.2 | | [46] | | | | |

Regarding the atomic electric polarizability $<\alpha_C>$, the three DFT methods yield highly similar results, with an average value of 7.96 atomic units (a.u.), representing a reduction of approximately 29.5% compared to the recommended value in [45]. This is noteworthy considering the utilization of an expanded basis set and a reliable functional. In contrast, the MP2/6-31G method shows a larger deviation, exhibiting a difference of 55.1% from the recommended value.

Concerning molecular polarizabilities, the predicted MP2 values are of the same order of magnitude, but are smaller compared with the DFT methods. According to Table 11, the PBE-D3 method consistently gives higher molecular polarizabilities than B3LYP, B3LYP-D3, and MP2 methods. Specifically, for the $D_{6d}$ isomer, both B3LYP and B3LYP-D3 produce very similar polarizabilities, averaging at 201.58 a.u., which is 7.9% lower than the PBE-D3 result. The dispersion correction in B3LYP only contributes minimally to the differences observed. On the other hand, the MP2/6-31G method yields a polarizability 6.1% smaller than the average of B3LYP and B3LYP-D3, and 13.6% smaller than PBE-D3. Interestingly, our MP2/6-31G polarizability is only 1.2% different from the one obtained in a previous study [45] using PBE/6-311+G(d). Moreover, an earlier research [43-44] reports an average polarizability of 215.60 a.u. obtained with the B3LYP method using 6-31+G(d) and 6-311++G(d,p) basis sets, showing an increase of 6.5% compared to the average of our B3LYP and B3LYP-D3 calculations with the cc-pVTZ basis set, and a 12.2% increase compared to our MP2/6-31G value. Regarding the $O_h$ isomer, the inclusion of dispersion corrections in the B3LYP method has a noticeable but relatively minor effect. The average polarizability calculated using B3LYP and B3LYP-D3 is 199.12, which is 1.86% lower than the polarizability obtained from PBE-D3. On the other hand, the MP2/6-31G method yields a polarizability that is 14.94% smaller than the PBE-D3 value and 13.3% smaller than the average polarizability obtained from the combination of B3LYP and B3LYP-D3. For the $D_{12h}$ isomer, we see a comparable trend to the previous isomer, where the inclusion of dispersion correction in B3LYP has minimal effect, resulting in B3LYP-D3 values closely resembling those of B3LYP. The average polarizability obtained from B3LYP and B3LYP-D3 is 498.43 a.u., indicating a decrease of 9.1% compared to the polarizability obtained from PBE-D3. In a similar manner, the $D_{6h}$ isomer's polarizability is highly similar for both the B3LYP and B3LYP-D3 methods due to the negligible impact of dispersion corrections. The average polarizability from these two methods is 259.1, representing a decrease of 3.8% compared to the polarizability obtained from PBE-D3. However, the MP2/6-31G method shows a polarizability that is 17.1% smaller than the PBE-D3 value and 13.8% smaller than the polarizability obtained from B3LYP and B3LYP-D3. Similar to $O_h$, the $D_{2d}$ isomer follows the same trend. The inclusion of dispersion correction in B3LYP has a small, yet significant, impact. The average polarizability obtained from B3LYP and B3LYP-D3 is 259.96, indicating a decrease of 1.98% compared to the polarizability obtained from PBE-D3. PBE-D3 and B3LYP-D3 methods exhibiting a consistent agreement regarding the ranking of polarizabilities in ascending order, with the exception of $D_{6h}$ and $D_{2d}$ isomers, which are ranked oppositely between the two methods. All the methods unanimously determine that $O_h$ ($D_{12h}$) possesses the smallest (highest) polarizability. Considering the minimal differences between the polarizabilities obtained from B3LYP and B3LYP-D3, we investigated the variations in polarizability within the isomer series when comparing PBE to the average of B3LYP and B3LYP-D3, noted as "Mean (B3LYP, B3LYP-D3)". The results, in ascending order, are 3.77 for $O_h$, 5.24 for $D_{2d}$, 10.29 for $D_{6h}$, 17.45 for $D_{6d}$, and 49.66 for $D_{12h}$. However, when comparing PBE to MP2, the polarizability differences are 29.69 for $D_{6d}$, 30.32 for $O_h$, and 46.12 for $D_{6h}$. There is no consistent correlation observed between the rankings of the smallest and largest differences in polarizability between PBE- Mean (B3LYP, B3LYP-D3) and PBE-MP2. It is worth noting that the polarizability of $D_{12h}$ is highly influenced by the choice of the DFT method, as it becomes similar to $D_{6h}$ when comparing DFT methods. Conversely, $O_h$ is the isomer that exhibits weak dependence on the DFT method, but it becomes comparable to $D_{6d}$ and shares similarities with it.

When comparing the sum of atomic Polarizabilities with the molecular polarizability, the two values are found to be comparable in the $O_h$ isomer when calculated within the MP2 framework. For the other isomers, the additive sum of atomic Polarizabilities differs much from the molecular value. Finally, compared to other more sophisticated methods, the predicted MP2 and DFT atomic polarizability values of carbon atom are significantly underestimated.

To summarize, as polarizability is a second-order property, it is highly sensitive to the molecular geometry and to the choice of wavefunction or functional used in calculations, as evident from its comparison with the electric dipole moment. The impact of dispersion terms is relatively minor in the B3LYP method, but it becomes more substantial in the PBE method. However, quantitatively assessing the individual contributions of the functional and dispersion corrections for PBE remains a challenging task.

### 3.5. Vibration Frequencies and Infrared Spectra

The $C_{24}$ isomers has a total of 66 vibrational modes (3N-6, where N is the number of atoms). Figure 5 displays the vibrational frequencies and infrared (IR) spectra of C24 isomers at the PBE-D3 level of theory using the cc-pVTZ basis set. The results of B3LYP/cc-pVTZ, B3LYP-D3/cc-pVTZ and MP2/6-31G are reported in Figures S1and S2 of the supplementary materials file. These figures visually represent the data and illustrate the variations in vibrational frequencies and IR spectra among the different isomers. The isomer with the highest symmetry, the ring isomer, exhibits the fewest peaks (2) in its spectrum. The $O_h$ isomer, with lower symmetry, has 3 peaks, followed by the Bracelet $D_{2d}$ isomer with 4 peaks, followed by the sheet $D_{6h}$ with 5 peaks, and finally the $D_{6d}$ with 13 peaks, according to the three DFT methods. Furthermore, the absorption wavelengths obtained from DFT calculations are quite comparable to those from MP2 calculations. The influence of electronic correlation and dispersion effects becomes evident when comparing the wavelengths originating from the MP2 calculation with those from the and PBE one. The way by which the electronic correlation is taken into account, as well as the consideration of non-covalent interactions, becomes evident when comparing the absorption wavelengths obtained from the PBE, B3LYP, B3LYP-D3 and MP2 methods.

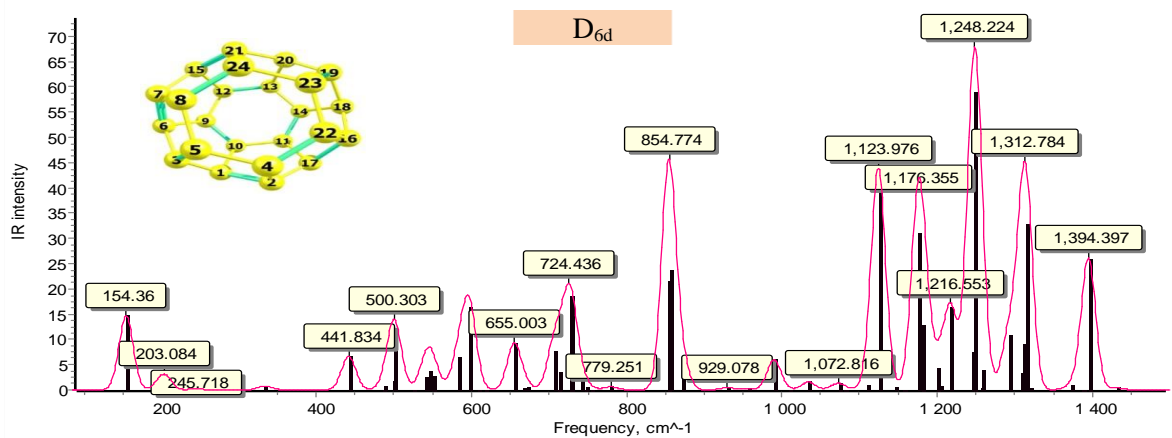

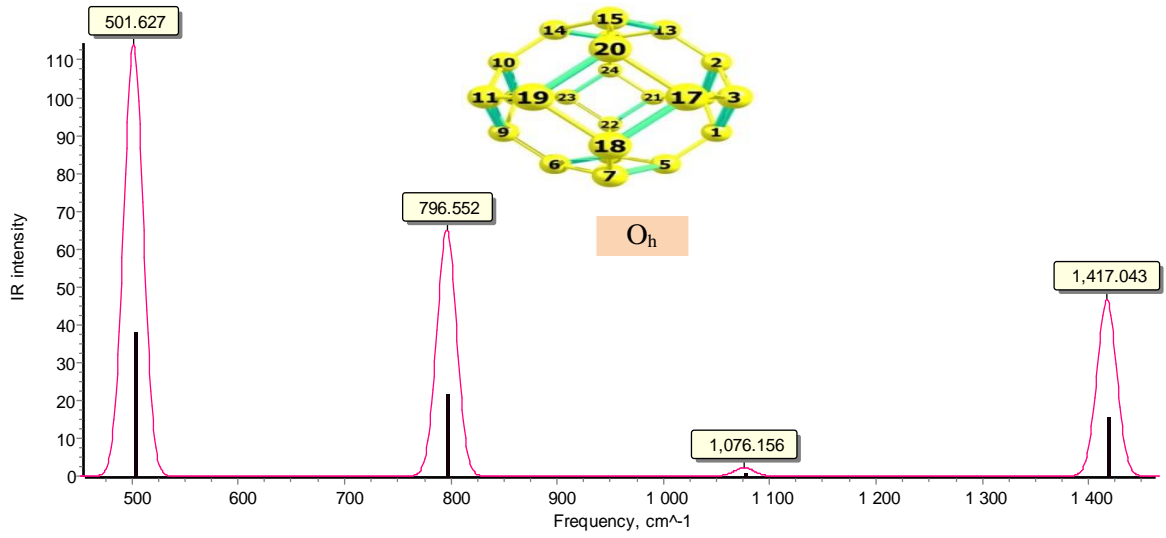

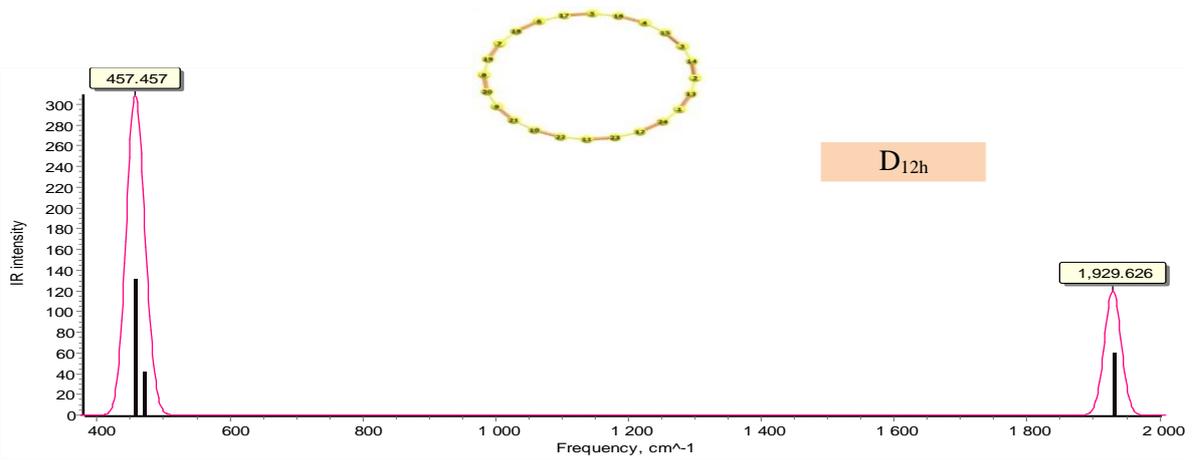

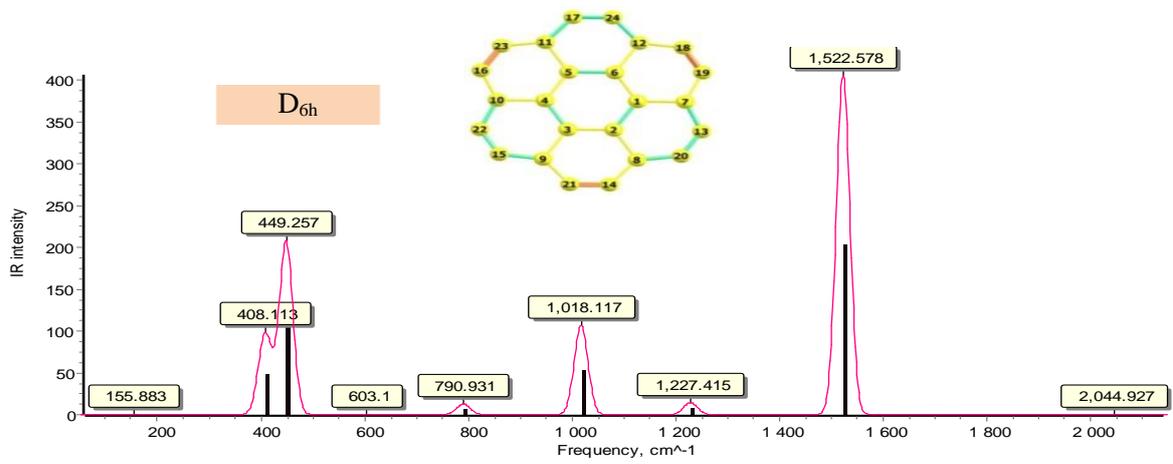

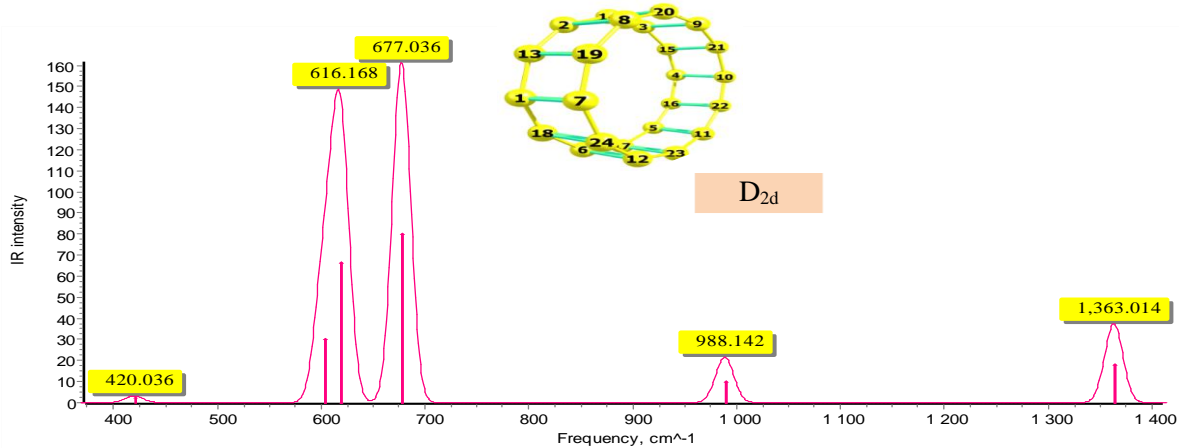

**Fig. 6.** Infrared spectra of the $C_{24}$ isomers at PBE-D3/cc-PVTZ level.

### 3.6. NMR Spectra for the $C_{24}$ isomers

We report in Tables S3, S4, S5, S6, S7, and S8 placed in the supplementary material file, the values (in ppm) of the isotropic and anisotropic nuclear screening constants ($\sigma$) of the $C_{24}$ isomers calculated at the PBE, B3LYP, B3LYP-D3 levels of theory using the cc-pVTZ basis set.

These tables show that, for the $D_{6d}$ isomer, the average ($\sigma$) of atoms belonging to the pentagonal shape is equal to 20.772 ppm, whereas this average value is 54.121 ppm for atoms in the hexagonal ring. These different values allow us to distinguish clearly between the two types of atoms. For the $O_h$ and $D_{12h}$ isomers, all carbon atoms are quasi-equivalent and have a mean isotropic ($\sigma$) part of about 30.431 ppm. For the $D_{6h}$ isomer, the value of ($\sigma$) is distributed in three classes. First, the mean value of ($\sigma$) for the central hexagonal cycle (atoms $C_1$-$C_6$) is about 29.784ppm. Second, for peripheral atoms, linked to the central hexagonal cycle ($C_7$-$C_{12}$), the mean value of ($\sigma$) is 67.948 ppm. Third, this average mean increases to 99.067ppm for peripheral atoms, not linked to the central hexagonal cycle ($C_{13}$-$C_{24}$), which indicates that the electron density on these atoms is higher, leading to a shielding effect. For the $D_{2d}$ isomer, the value of ($\sigma$) is distributed in two classes. First, the mean value of ($\sigma$) (atoms $C_3, C_6, C_9, C_{12}, C_{13}, C_{16}, C_{19}$, and $C_{22}$) is about 29.784ppm. Second, for peripheral atoms, linked to the central hexagonal cycle ($C_7$-$C_{12}$), the mean value of ($\sigma$) is 67.948 ppm. As for the remaining atoms, the mean value of ($\sigma$) is 59.882ppm.

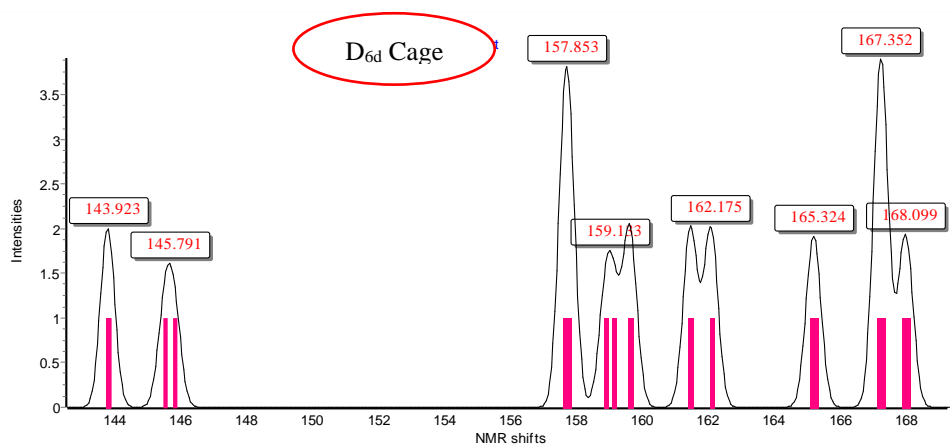

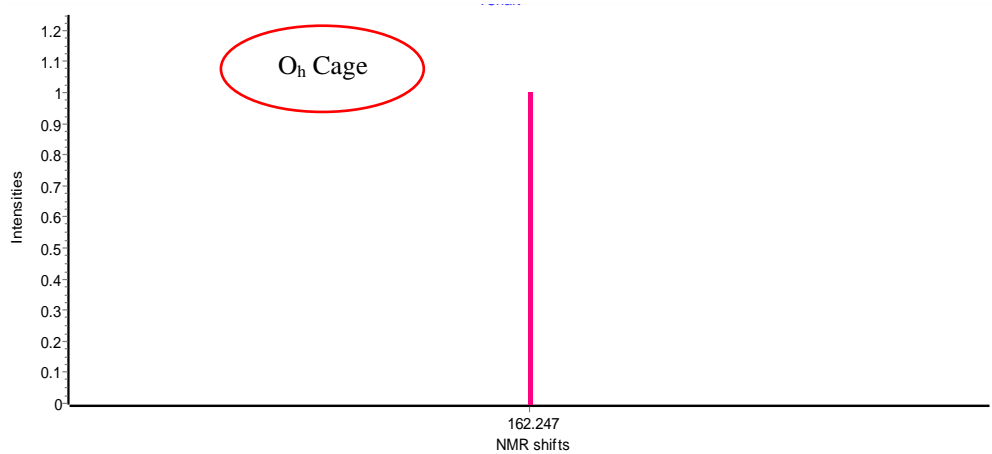

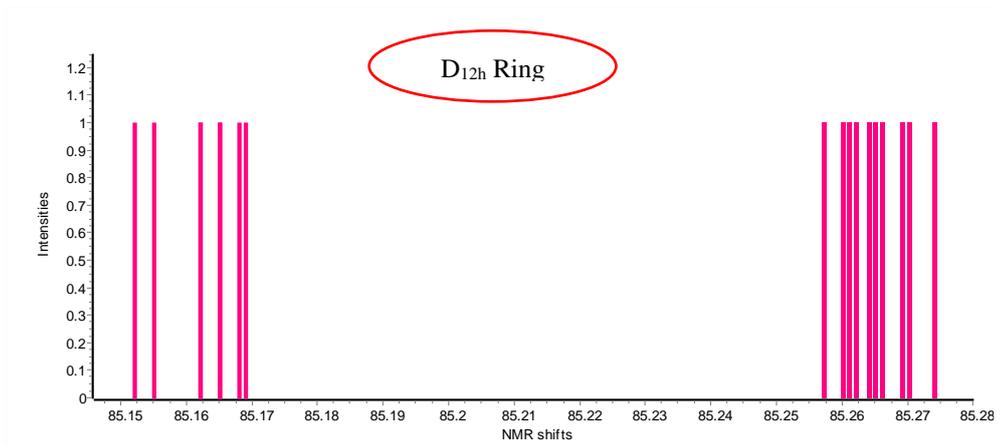

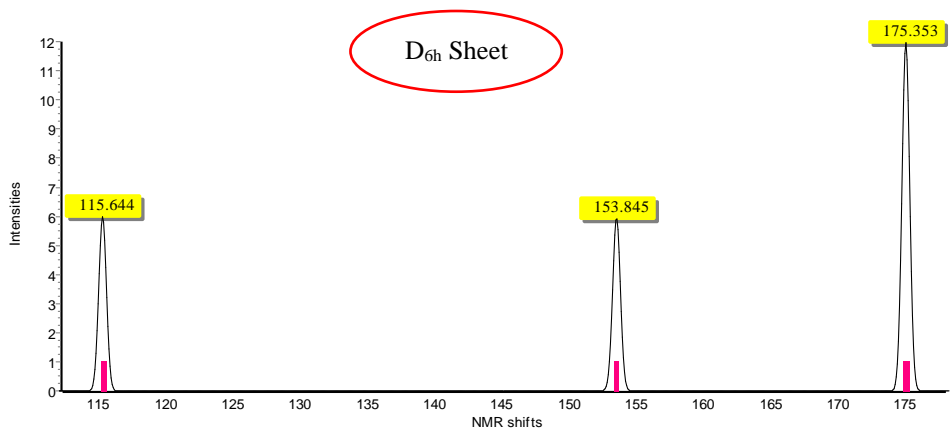

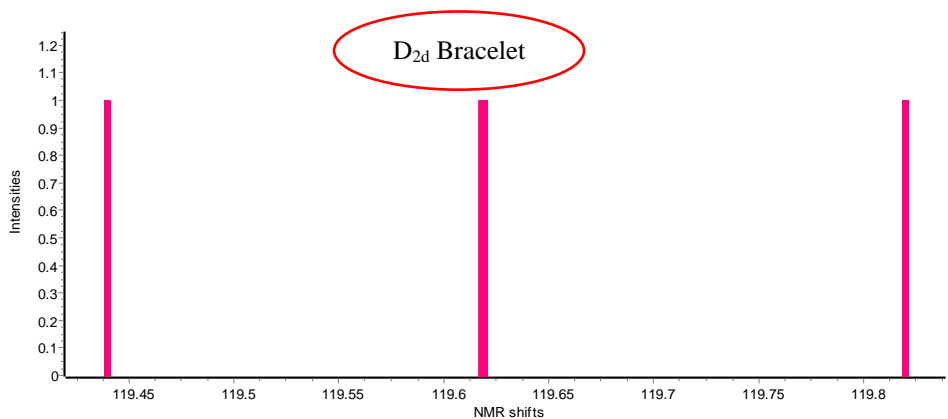

**Fig. 7.** NMR spectra of the $C_{24}$ isomers at PBE-D3/cc-PVTZ level.

## 5. Conclusion

In this study, we investigated multiple aspects of $C_{24}$ isomers ($D_{6d}$, $O_h$, $D_{12h}$, $D_{6h}$, and $D_{2d}$) using the PBE-D3, B3LYP, B3LYP-D3 (cc-pVTZ) and MP2/6-31G methods. We analyzed geometry optimizations, relative stability, static electric polarizabilities, nuclear screening constants, gap and atomization energies, thermodynamic analysis, vibration frequencies, as well as infrared and NMR spectra.

Consistently, the $D_{6h}$ isomer was found to be the most stable among the studied $C_{24}$ isomers, while the newly proposed $D_{2d}$ isomer exhibited the least stability. The B3LYP and B3LYP-D3 methods consistently predicted higher energy gap (GE) values compared to PBE, with an average DFT (PBE, B3LYP and B3LYP-D3) GE of 1.89 eV. In contrast, the MP2 method indicated a significantly higher GE value of 7.63 eV, representing a substantial increase of approximately 75%. Regarding the electronic character, PBE classified the $D_{12h}$, $D_{6d}$, $D_{2d}$, and $O_h$ isomers as semiconductors, whereas B3LYP or B3LYP-D3 classified them as weak insulators. On the other hand, the MP2 method predicted a distinct insulating character for the $D_{6h}$, $O_h$, and $D_{6d}$ isomers, with the $D_{6h}$ isomer being the most insulating among the studied isomers. The PBE-D3 method consistently provides higher polarizabilities of the $C_{24}$ isomers compared to B3LYP, B3LYP-D3 and MP2 methods. This underscores the significance of considering electronic correlation and dispersion effects for accurate prediction of polarizabilities.

The results obtained from different methods emphasize the impact of methodology on the predicted properties, highlighting the need for careful analysis and interpretation of theoretical results for diverse geometries. Notably, the study reveals that $O_h$ has the smallest polarizability, and $D_{12h}$ exhibits the highest polarizability, indicating a significant influence of the applied electric field on its electronic cloud, a consensus among all methods. Moreover, the polarizability of $D_{12h}$ significantly depends on the chosen DFT method, while $O_h$ shows less sensitivity but shares similarities with $D_{6d}$. Additionally, the $D_{12h}$ isomer displayed the highest number of absorption peaks, while the calculated nuclear screening constants allowed for distinguishing the magnetically equivalent carbon atom groups in each isomer.

In conclusion, our study provides valuable insights into the stability, electronic properties, and polarization behavior of $C_{24}$ isomers. The findings underscore the importance of methodology selection and electronic correlation and dispersion considerations in accurately predicting properties and interpreting theoretical results for the $C_{24}$ isomers with varying geometries.

## Data Availability Statement
Not applicable.

## Declaration of competing interest
The authors report no conflicts of interest. The authors alone are responsible for the content and writing of this article.

## Acknowledgments
N. Chamoun acknowledges support from the PIFI-CAS program, from Humboldt Foundation, and from ICTP Associate program.

**Supporting Information Available:** Are reported as supplemental information: Cartesian coordinates of C24-D2d isomer optimized at the PBE-D3/cc-pVTZ level of theory, optimized energies and thermodynamic analysis using the four methods PBE, B3LYP, B3LYP-D3 and MP2, nuclear screening constants using PBE, B3LYP and B3LYP-D3 and IR spectra paraphs using MP2, B3LYP and B3LYP-D3. This material is available free of charge via the Internet at: